\documentclass[aps,letterpaper,10pt,twocolumn,floats,showpacs,amsmath,amsfonts,amssymb,pre]{revtex4-1}

\usepackage{graphicx}
\usepackage[colorlinks=true,citecolor=blue]{hyperref}
\usepackage[caption=false]{subfig}
\usepackage{pstricks}
\usepackage{amsmath}
\usepackage{amsbsy}
\usepackage{verbatim}
\begin{document}

\preprint{APS/123-QED}

\title{Autonomous quantum Maxwell's demon based on two exchange-coupled quantum dots}

\author{Krzysztof Ptaszy\'{n}ski}
 \email{krzysztof.ptaszynski@ifmpan.poznan.pl}
\affiliation{%
 Institute of Molecular Physics, Polish Academy of Sciences, ul. M. Smoluchowskiego 17, 60-179 Pozna\'{n}, Poland
}%

\date{\today}

\begin{abstract}
I study an autonomous quantum Maxwell's demon based on two exchange-coupled quantum dots attached to the spin-polarized leads. The principle of operation of the demon is based on the coherent oscillations between the spin states of the system which act as a quantum iSWAP gate. Due to the operation of the iSWAP gate one of the dots acts as a feedback controller which blocks the transport with the bias in the other dot, thus inducing the electron pumping against the bias; this leads to the locally negative entropy production. Operation of the demon is associated with the information transfer between the dots, which is studied quantitatively by mapping the analyzed setup onto the thermodynamically equivalent auxiliary system. The calculated entropy production in a single subsystem and information flow between the subsystems are shown to obey a local form of the second law of thermodynamics, similar to the one previously derived for classical bipartite systems.

\end{abstract}

\maketitle


\section{\label{sec:intro}Introduction}
Maxwell's demons, i.e. physical system in which the feedback control may lead to the locally negative entropy production, have become a standard example of the relation between thermodynamics and information theory~\cite{maruyama2009, parrondo2015}. A convenient way to experimentally realize such systems is provided by electronic circuits~\cite{koski2016}. While in original thought experiment of Maxwell the principle of operation of the demon has been based on the action of some external intelligent agent, nowadays it is recognized that the locally negative entropy production may arise due to the information transfer between two coupled stochastic systems~\cite{strasberg2013, horowitz2014}. Such setups are referred to as autonomous Maxwell's demons. The physical example of such a system is a device consisting of two capacitively coupled quantum dots, in which the operation of the Maxwell's demon has been first studied theoretically by Strasberg \textit{et al.}~\cite{strasberg2013}, and later experimentally by Koski \textit{et al.}~\cite{koski2015}. Horowitz and Esposito~\cite{horowitz2014} have proposed a consistent approach, based on the formalism of the stochastic thermodynamics~\cite{seifert2012}, to describe the information flow within discrete Markovian networks; this allows to analyze the operation of the autonomous Maxwell's demons quantitatively. However, this formalism is confined to the description of the certain class of classical stochastic systems, referred to as the bipartite systems, i.e. ones in which dynamics of a single subsystem depends on the state of the other, but there are no transitions inducing the simultaneous change of both subsystems. In particular, this approach cannot be directly applied to the quantum coherent systems. 

While Maxwell's demons in the quantum coherent systems have been already studied both theoretically~\cite{lloyd1997, kim2011, park2013, pekola2016, elouard2017} and experimentally~\cite{camati2016, cottet2017}, the analysis has been mainly confined to the non-autonomous setups, i.e. requiring the external feedback control. As an exception, Champman and Miyake~\cite{chapman2015} have considered an autonomous demon in which the external control has been based on the cyclic interaction with a tape of memory qubits operating at the periodic steady state. Translation of the memory tape has been assumed to be deterministic. In contrast, here I analyze the system in which the principle of operation is based on the quantum information exchange between two coherently interacting quantum stochastic systems operating at the steady state of the time-independent evolution generator. In this way, the considered setup does not require any deterministic time-dependent driving, in a direct analogy to the classical autonomous demons analyzed in Refs.~\cite{strasberg2013, horowitz2014, koski2015}.

Specifically, the studied system is based on two exchange-coupled quantum dots attached to the spin-polarized leads. The principle of operation is based on the coherent oscillations between the spin states of the system, which can be interpreted as an operation of the iSWAP gate between the spin qubits inducing the information flow between the quantum dots. Although the system is not a bipartite one, its dynamics can be mapped onto the auxiliary quantum model, which enables to separate contributions to the rate of change of quantum mutual information associated with dynamics of different subsystems. This demonstrates the possibility of the quantitative study of the information flow within the stochastic quantum systems. Moreover, it is shown that the sum of the entropy production in a single subsystem and the information flow from this subsystem to another one is always nonnegative, and thus obey a local version of the second law of thermodynamics, similar to the one derived by Horowitz and Esposito~\cite{horowitz2014} for classical bipartite systems.

The paper is organized as follows. Section~\ref{sec:model} presents the model of the considered double-dot system, as well as the master equation describing its dynamics. In Sec.~\ref{sec:results} I present the calculated thermodynamic quantities and discuss the results; this section contains also the description of the mapping procedure. Finally, Sec.~\ref{sec:conclusions} brings conclusions following from my results. The Appendix contains discussion of the energy exchange between the quantum dots.

\section{\label{sec:model}Model and methods}
\begin{figure}
	\centering
	\subfloat[]{\includegraphics[width=0.53\linewidth]{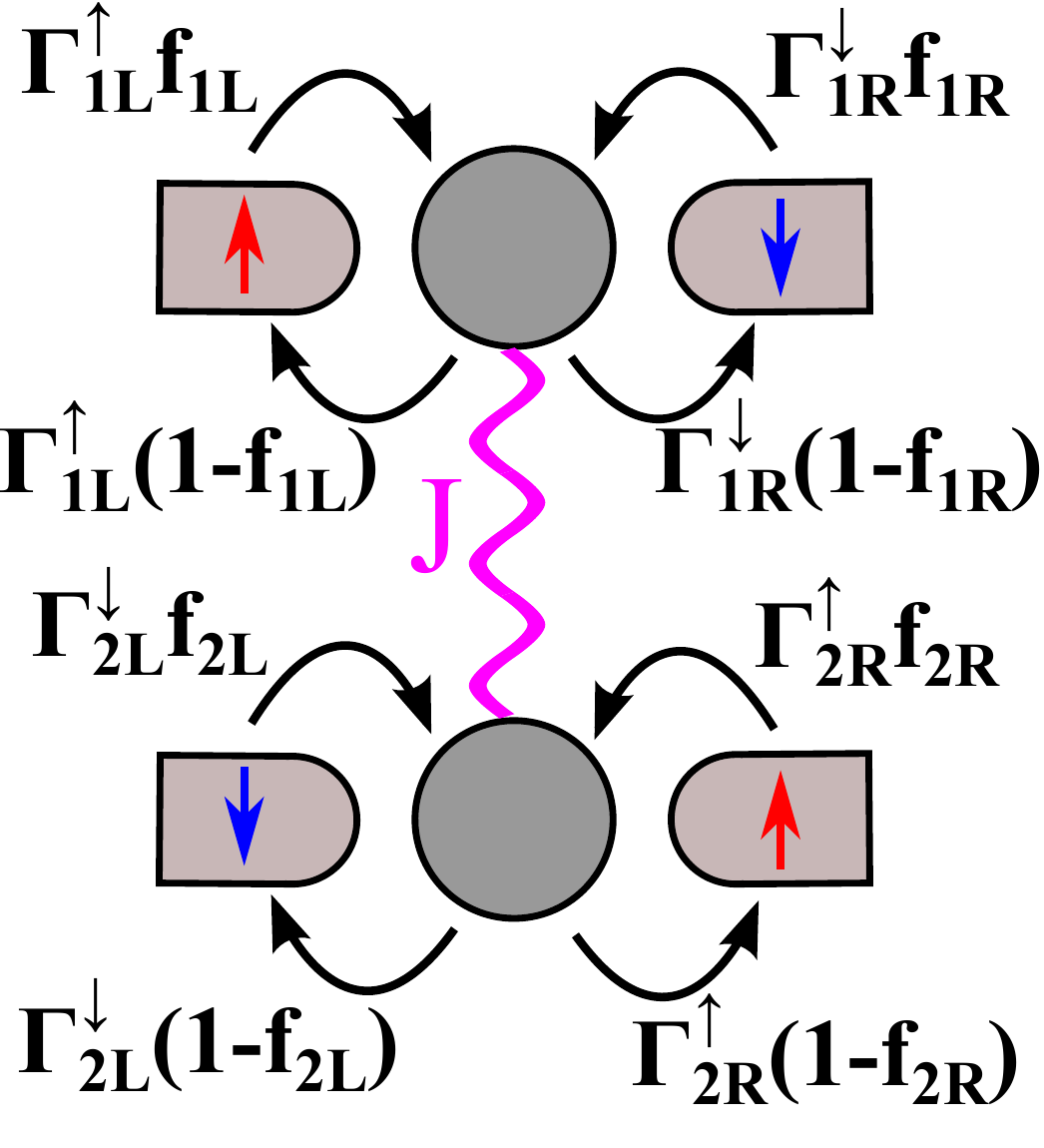}}
	\hspace{1.5mm}
	\subfloat[]{\includegraphics[width=0.44\linewidth]{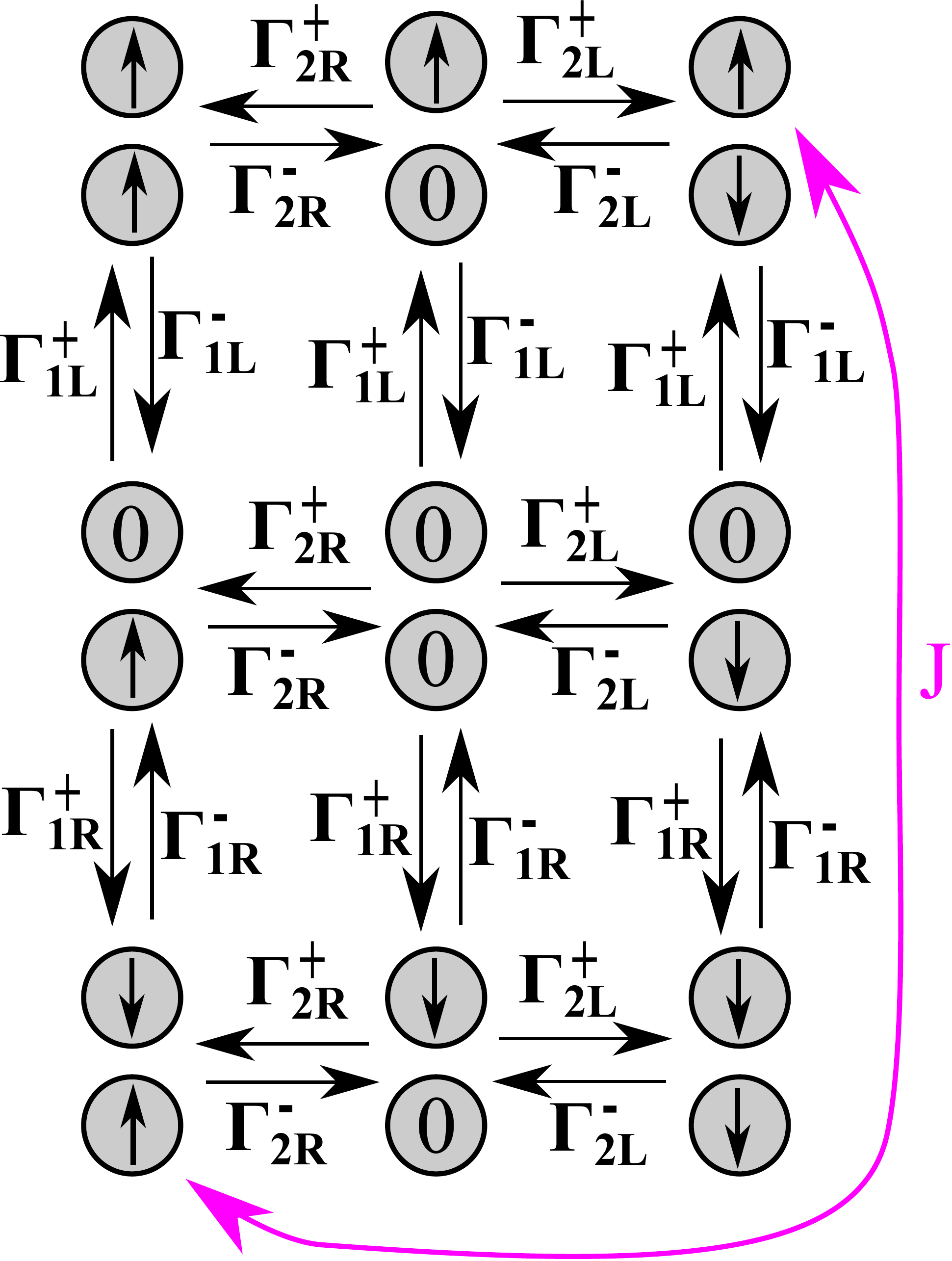}}
	\caption{a) Scheme of the studied system consisting of two exchange-coupled quantum dots attached to the spin-polarized leads. $J$ denotes the exchange interaction, $\Gamma_{i \nu}^\sigma$ the tunneling rates and $f_{i \nu}$ the Fermi distribution functions. b) Scheme of the dynamics of the system. Black arrows denote transitions associated with the electrons tunneling while the violet double-ended arrow denotes the coherent oscillations between the states ${|\! \! \uparrow \downarrow \rangle}$ and ${|\! \! \downarrow \uparrow \rangle}$. Notation as at the end of Sec.~\ref{sec:model}.}
	\label{fig:system}
\end{figure}

The analyzed system consists of two exchange-coupled single-level quantum dots, each weakly attached to two fully and collinearly spin-polarized leads, arranged in the anti-parallel way [Fig.~\ref{fig:system}~(a)]. The Hamiltonian of the isolated double dot-system (uncoupled to the leads) reads
 \begin{align} \label{hamiltonian}
  \hat{H}_D &=\sum_{i \in \{1,2\}} \sum_{\sigma \in \{\uparrow, \downarrow\}} \epsilon_i c_{i\sigma}^\dagger c_{i\sigma} + \sum_{i \in \{1,2\}} U_i n_{i \uparrow} n_{i \downarrow} \\ \nonumber &+J(\hat{S}_1^x \hat{S}_2^x+\hat{S}_1^y \hat{S}_2^y),
  \end{align} 

where $\epsilon_i$ is the orbital energy of the $i$th dot, $c_{i\sigma}^\dagger$ ($c_{i\sigma}$) is the creation (annihilation) operator of the electron with a spin $\sigma$ in the $i$th dot, $U_i$ is the intra-dot Coulomb interaction energy in the $i$th dot, $J$ is the exchange coupling between the dots and $S_i^\eta$ is the operator of the spin $\eta$-projection in the $i$th dot. The model assumes presence of the XY-type exchange coupling between the dots without electron tunneling between them. It was proposed theoretically that such a coupling can be obtained using the photon~\cite{trif2008} or phonon~\cite{wang2015} mediated interaction between the dots. The XY-type exchange interaction naturally generates the iSWAP quantum gate~\cite{wang2015, schuch2003}. Similar dynamics can be obtained using the more standard Heisenberg type coupling, but not using the Ising type coupling. I also assume that the intra-dot Coulomb interaction is large, such that double occupancy of the dot is not enabled (the strong Coulomb blockade regime). The states of the system can be then expressed in the basis of nine localized states: $\{{|00 \rangle}, {|0 \! \! \uparrow \rangle}, {|0 \! \! \downarrow \rangle}, {|\! \! \uparrow \! \!  0 \rangle}, {|\! \! \uparrow \uparrow \rangle}, {|\! \! \uparrow \downarrow \rangle}, {|\! \!  \downarrow \! \! 0 \rangle}, {|\! \! \downarrow \uparrow \rangle}, {|\! \! \downarrow \downarrow \rangle\}}$, where the first/second position in the ket corresponds to the first/second dot, $0$ denotes the empty dot and arrows denote the spin polarization of the electron occupying the dot.

Each dot is attached to two leads denoted as $i\nu$, where $i \in \{1,2\}$ denotes the dot to which the electrode is coupled, whereas $\nu=L$ ($\nu=R$) denotes the left (right) lead. The electrochemical potential and the temperature of the lead $i \nu$ are denoted as $\mu_{i \nu}$ and $T_{i \nu}$. I assume that $|J| \ll k_B T_{i\nu}, |\epsilon_i-\mu_{i\nu}|$, such that level splitting due to the exchange coupling is much smaller than the other energy scales, and therefore it does not influence the tunneling between the dots and the leads. This also provides, that the energy transfer between the dots can be neglected; the validity of this assumption is demonstrated in the Appendix. I also focus on the weak coupling regime, in which the level broadening due to lead-dot coupling can be neglected, i.e. $\hbar \Gamma_{i \nu}^\sigma \ll k_B T_{i\nu}, |\epsilon_i-\mu_{i\nu}|$, where $\Gamma_{i \nu}^\sigma$ is the tunneling rate between the $i$th dot and the lead $i \nu$ for a spin $\sigma$ \cite{li2005}. The spin dependence of the tunneling rates may result from different density of states for different spins in the leads~\cite{rudzinski2001, braun2004}. Transport can be then described by the master equation written in the Lindblad form~\cite{benenti2009, busl2010, luczak2014}:
\begin{align} \label{lindblad} \nonumber
\frac{d \hat{\rho}}{dt} = &-i \left[ \hat{H}_D, \hat{\rho} \right] \\ &+ \sum_{i, \nu, \sigma} \frac{\Gamma_{i\nu}^\sigma f_{i\nu}}{2} \left(2 c^{\dagger}_{i \sigma} \hat{\rho} c_{i \sigma}-c_{i \sigma} c^{\dagger}_{i \sigma} \hat{\rho} -\hat{\rho} c_{i \sigma} c^{\dagger}_{i \sigma} \right) \\ \nonumber
&+ \sum_{i, \nu, \sigma} \frac{\Gamma_{i \nu}^\sigma (1-f_{i\nu})}{2} \left(2 c_{i \sigma} \hat{\rho} c_{i \sigma}^{\dagger}-c_{i \sigma}^{\dagger} c_{i \sigma} \hat{\rho} -\hat{\rho} c_{i \sigma}^{\dagger} c_{i \sigma} \right), 
\end{align}
in which for simplicity $\hbar=1$ is taken. The first term of the right-hand side of the equation describes the coherent evolution of the density matrix of the system $\hat{\rho}$ associated with the oscillations between the spin states due to the presence of the exchange coupling, whereas the next two terms describe the sequential tunneling of electrons between the dots and the leads (to the dot or from the dot, respectively). Here $f_{i \nu}=f[(\epsilon_i-\mu_{i\nu})/k_B T_{i\nu}]$ is the Fermi distribution function of the electrons in the lead $i \nu$. The used master equation in the high voltage limit corresponds to the one derived by Gurvitz and Prager~\cite{gurvitz1996, gurvitz1998}. In contrast to the Pauli master equation for populations in the eigenstate basis~\cite{breuer2002} (also referred to as the diagonalized master equation~\cite{poltl2009}), often used in the case of finite voltages, it takes into account the coherent oscillations between the spin states. 

In the following part of the paper all temperatures are assumed to be equal: $T_{i \nu}=T$. Moreover, I assume that the leads are fully spin-polarized such that $\Gamma_{1L}^\downarrow=\Gamma_{1R}^\uparrow=\Gamma_{2L}^\uparrow=\Gamma_{2R}^\downarrow=0$. The following notation will be also sometimes used: $\Gamma_{i \nu}^+=\Gamma_{i \nu} f_{i \nu}$ and $\Gamma_{i \nu}^-=\Gamma_{i \nu} (1-f_{i \nu})$, with $\Gamma_{1L}=\Gamma_{1L}^\uparrow$, $\Gamma_{1R}=\Gamma_{1R}^\downarrow$, $\Gamma_{2L}=\Gamma_{2L}^\downarrow$, and $\Gamma_{2R}=\Gamma_{2R}^\uparrow$. Dynamics of the system can be illustrated using the graphical scheme shown in Fig.~\ref{fig:system}~(b).

\section{\label{sec:results} Results}
\subsection{\label{subsec:entropy} Entropy production}
Now I analyze the thermodynamic flows in the considered systems. The study is confined to the analysis of the steady state. The particle currents through the first and the second dot (from the left to the right lead), denoted as $\dot{N}_1$ and $\dot{N}_2$, can be evaluated using the rate equation formalism~\cite{nazarov2009}. They are given by the following expressions:
\begin{align} \label{partcur1}
\dot{N}_1 & = \Gamma_{1L}^\uparrow \sum_{\lambda} \left[f_{1L} p_{0\lambda}-(1-f_{1L}) p_{\uparrow \lambda }\right], \\ \label{partcur2}
\dot{N}_2 & = \Gamma_{2L}^\downarrow \sum_{\kappa} \left[f_{2 L} p_{\kappa 0}-(1-f_{2 L}) p_{\kappa \downarrow}\right],
\end{align}
where $p_{\kappa \lambda}$ is the steady state probability of the state $|\kappa \lambda \rangle$, with $\kappa, \lambda \in \{0, \uparrow, \downarrow\}$, calculated using Eq.~\eqref{lindblad} by taking $d\hat{\rho}/dt=0$. Here, due to the full and anti-parallel spin polarization of the leads, transport is enabled only when the exchange coupling $J$ is non-zero and thus the oscillations between the states ${|\! \! \uparrow \downarrow \rangle}$ and ${|\! \! \downarrow \uparrow \rangle}$ take place; because each spin-flip in the one dot is associated with the spin-flip in the other dot, steady state particle currents in both dots are equal ($\dot{N}_1=\dot{N}_2$). However, the other thermodynamics currents (energy, heat and entropy flows) are in general non-equal due to the difference of the voltages. Since the energy exchange between the dots can be neglected (due to $|J| \ll |\epsilon_i-\mu_{i \nu}|, k_B T$; see Appendix for details), the entropy production rate in a single dot is fully determined by the particle tunneling. Using the definition of the entropy change $\Delta S=Q/T$ and standard formula for the Joule heating $\dot{Q}_i=\dot{N}_i V_i$, where $\dot{Q}_i$ is the rate of heat generation due to the electron tunneling through the $i$th dot and $V_i=\mu_{iL}-\mu_{iR}$ is the voltage applied to the $i$th dot, one obtains the entropy production rate in the $i$th dot:
\begin{align} \label{entrcur}
\dot{\sigma}_i=\frac{\dot{Q}_i}{T}=\frac{\dot{N}_i V_i}{T}.
\end{align}

%
\begin{figure}
	\centering
	  	\includegraphics[width=0.9\linewidth]{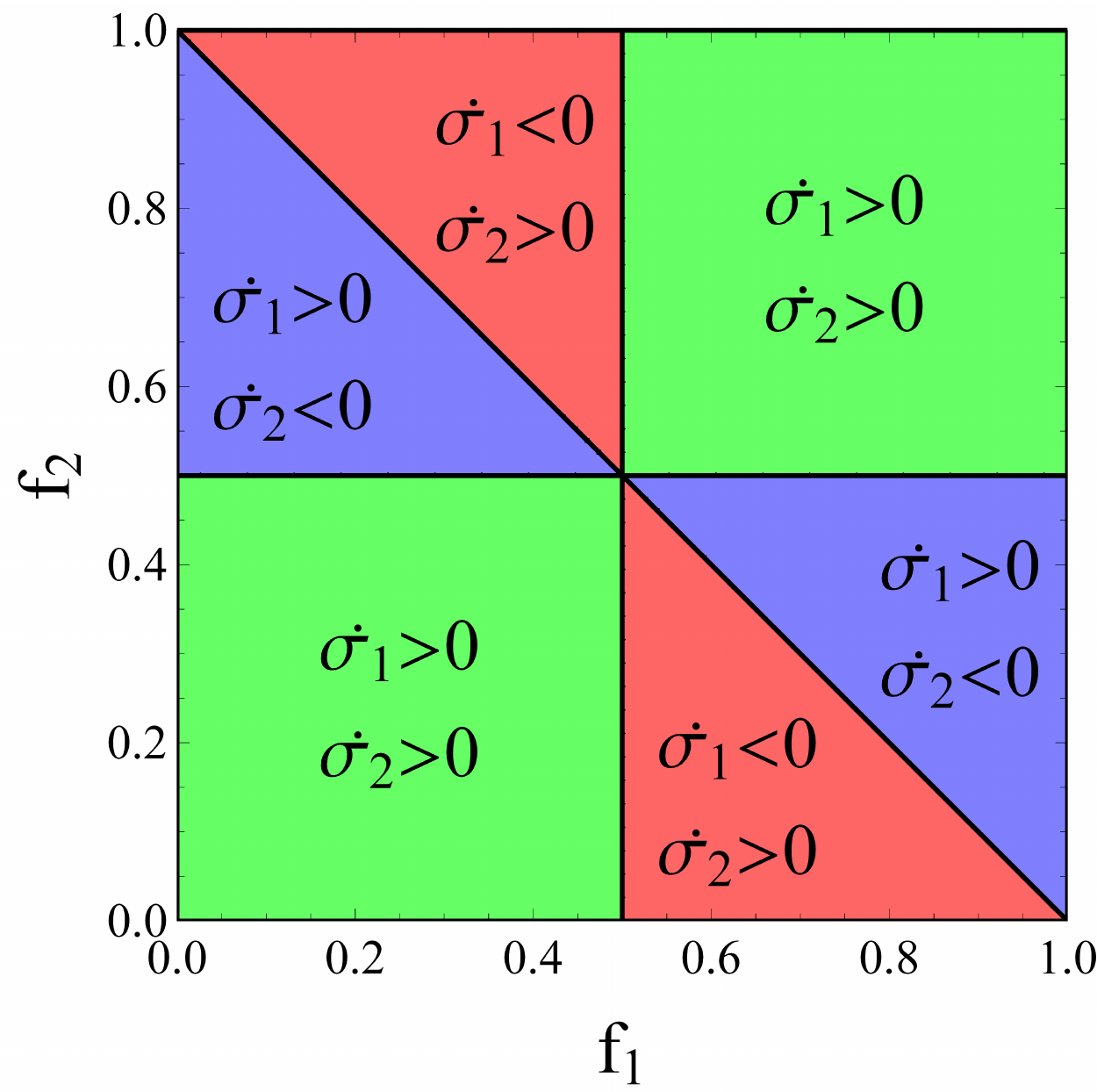}		
	\caption{Phase diagram of the system over $f_1=f(V_1/k_B T)$ and $f_2=f(V_2/k_B T)$ for arbitrary nonzero values of $\Gamma_{1L}^\uparrow$, $\Gamma_{1R}^\downarrow$, $\Gamma_{2L}^\downarrow$, $\Gamma_{2R}^\uparrow$, $\epsilon_i$ and $J$ showing the different operation modes: blue and red regions correspond to the Maxwell demon modes, with red (blue) regions corresponding to the negative entropy production in the first (second) dot; the green region corresponds to the mode in which entropy production in both dots is positive.}
	\label{fig:phase}
\end{figure}
%

It appears, that for the oppositely polarized voltages (i.e. $V_1 V_2<0$) the current in one dot flows against the bias (except the case of $V_1=-V_2$ when $\dot{N}_1=\dot{N}_2=0$), which (taking into account that energy transfer between the dots can be neglected) results in the negative entropy production in one of the dots; as Fig.~\ref{fig:phase} shows, the sign of the entropy production in both dots depends only on the parameters $f_1=f(V_1/k_B T)$ and $f_2=f(V_2/k_B T)$, which is a result of the equality of the particle currents flowing through the dots. Thus, the system acts as an autonomous Maxwell's demon. This behavior is the result of the coherent oscillations between the states ${|\! \! \uparrow \downarrow \rangle}$ and ${|\! \! \downarrow \uparrow \rangle}$, which can be described as an operation of the quantum iSWAP gate. Similar current flow against the bias induced by the operation of the SWAP gate has been already studied by Strasberg \textit{et al.}~\cite{strasberg2014}, however in a non-autonomous setup. 

Let us now analyze the principle of operation of the demon in detail. Without loss of generality, I focus on situation when the voltage bias in the first dot is positive and high, and thus $f_1$ is close to 0. Since the right lead is spin-down polarized, the accumulation of the spin $\uparrow$ electrons in the first dot takes place. Let us now assume, that the voltage bias in the second dot is negative and has a smaller value. Transport with the bias would require the spin-flip $\uparrow \rightarrow \downarrow$ in the second dot induced by the operation of the iSWAP gate generating the transition ${|\! \! \downarrow \uparrow \rangle} \rightarrow {|\! \! \uparrow \downarrow \rangle}$; but, due to the accumulation of the spin $\uparrow$ electrons in the first dot, probability of such a process is low. On the other hand, thermally excited tunneling against the bias is enabled by the operation of the iSWAP gate, which induces the spin-flip $\downarrow \rightarrow \uparrow$ in the second dot with a relatively high probability. In this way, the first dot acts as a feedback controller which reads out the spin state of the electron in the second dot and blocks the tunneling with the bias, while enabling transport in the reverse direction; this leads to the electron pumping against the bias. After each such process, spin $\downarrow$ can tunnel out from the first dot to the right lead and is replaced by a spin $\uparrow$ electron from the left lead. This can be interpreted as a resetting of the memory of the demon.

\subsection{\label{subsec:mapping} Mapping onto the auxiliary systems and the information flows}

In the autonomous Maxwell's demons the negative entropy production in one subsystem is enabled due to the information flow to the other subsystem. Let us now describe this flow quantitatively. As previously mentioned, the autonomous quantum dot demons studied before, based on the Coulomb coupling between dots, were bipartite systems; in such systems the information flow can be described using the approach of Horowitz and Esposito~\cite{horowitz2014}. Specifically, the bipartite scheme enables to separate the contributions to the rate of change of mutual information associated with the dynamics of the first and the second subsystem in a rigorous way. The system analyzed now is not a bipartite one because the coherent oscillations induce the simultaneous changes in both the first and the second dot. It is, therefore, not obvious how to separate the contributions to the mutual information rate associated with different subsystems. However, I show that the information flow can be calculated by mapping the system onto the thermodynamically equivalent auxiliary one, which can be interpreted as a bipartite system. This mapping involves formal duplication of the system in a way similar to presented by Barato and Seifert~\cite{barato2014} for the classical systems.

%
\begin{figure}
	\centering
	\subfloat[]{\includegraphics[width=0.94\linewidth]{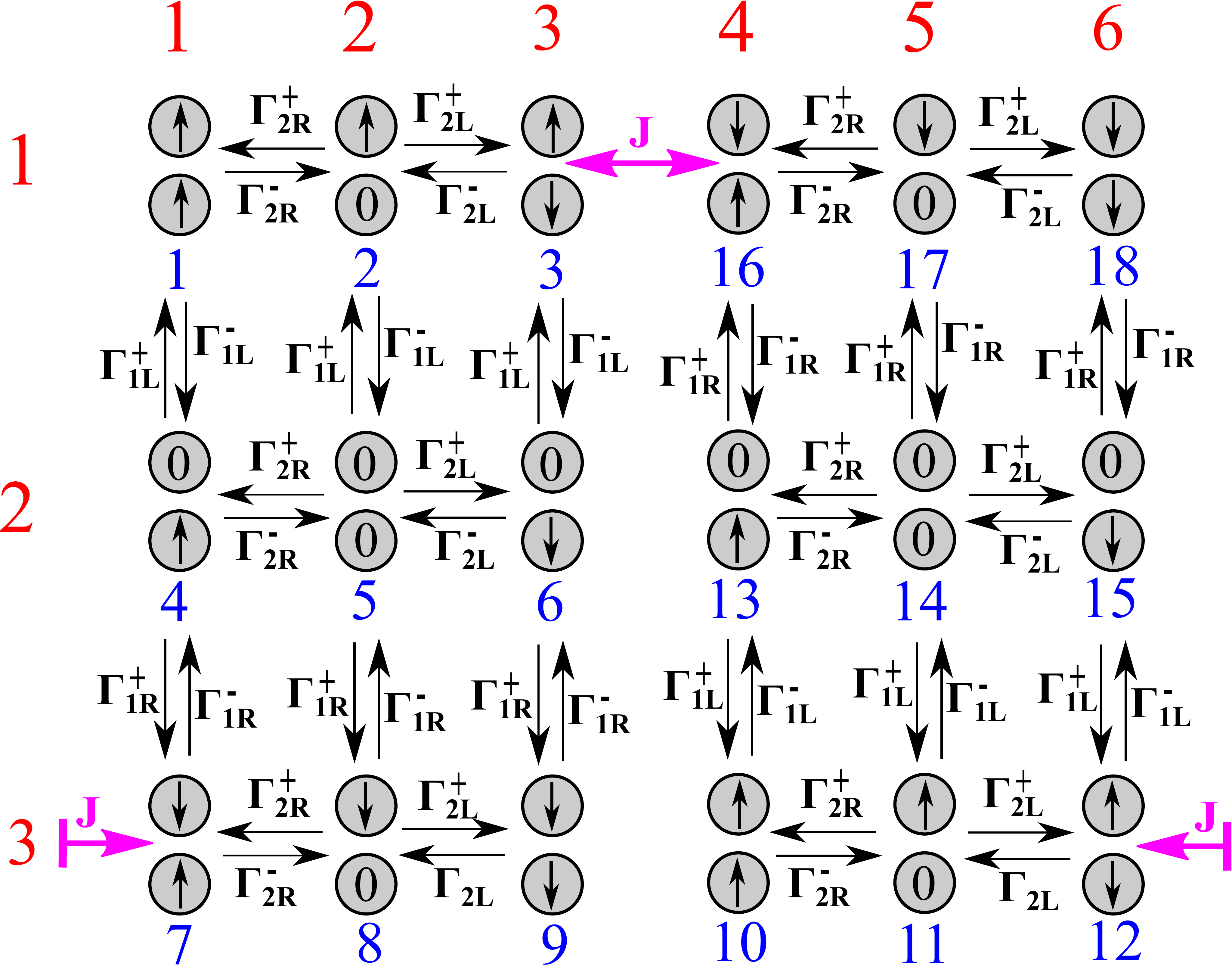}} \\
	\subfloat[]{\includegraphics[width=0.94\linewidth]{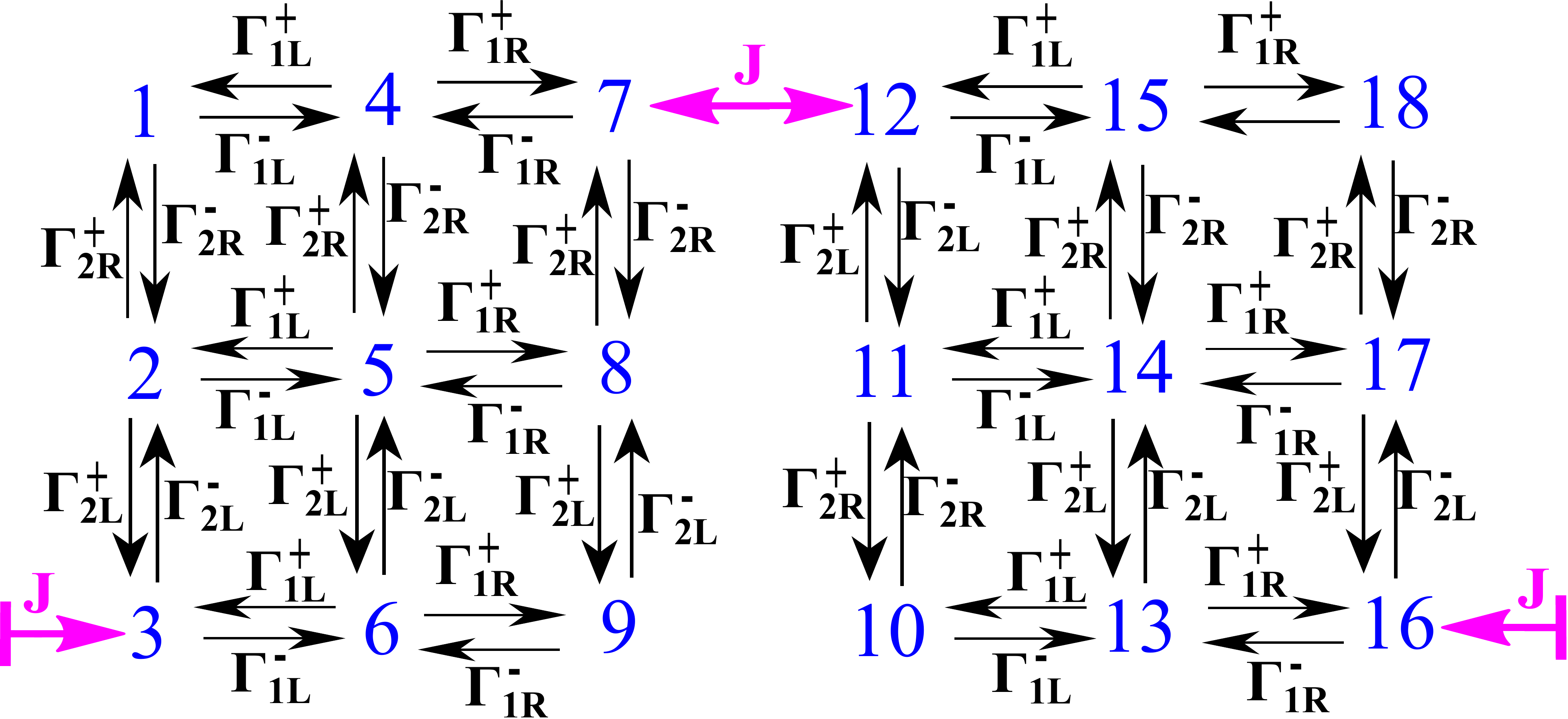}}
	\caption{(a) Scheme of the dynamics of the auxiliary system $A$. Notation of the tunneling rates defined at the end of Sec.~\ref{sec:model}. Blue numbers below the circles representing quantum dots denote the states $|i \rangle$ of the auxiliary system. Red numbers on the left and top side denote rows and columns. (b) Alternative arrangement with transitions in the second dot placed in columns, referred to as the auxiliary system $B$.}
	\label{fig:map}
\end{figure}
%

Let us use the following notation of the basis states: ${|\! \! \uparrow \uparrow \rangle}=|1 \rangle$, ${|\! \! \uparrow \! \!  0 \rangle}=|2 \rangle$, ${|\! \! \uparrow \downarrow \rangle}=|3 \rangle$, ${|0 \! \! \uparrow \rangle} = |4 \rangle$, ${|00 \rangle}=|5 \rangle$, ${|0 \! \! \downarrow \rangle}=|6 \rangle$, ${|\! \! \downarrow \uparrow \rangle}=|7 \rangle$, ${|\! \!  \downarrow \! \! 0 \rangle}=|8 \rangle$, ${|\! \! \downarrow \downarrow \rangle}=|9 \rangle$. Elements of the density matrix of the studied system are denoted as $\rho_{i,j}=\langle i | \hat{\rho}| j \rangle$. One can note that the coherent oscillations between the spin states are associated with the dynamics of the density matrix elements $\rho_{3,3}$, $\rho_{7,7}$ and $\rho_{3,7}=\rho_{7,3}^*$:
\begin{align} \label{dynorig}
&\dot{\rho}_{3,3}=\Gamma_{2L}^+ \rho_{2,2}-\gamma_{L}^{-} \rho_{3,3}+\Gamma_{1L}^+ \rho_{6,6}-J \text{Im}(\rho_{3,7}), \nonumber \\
&\dot{\rho}_{7,7} =\Gamma_{1R}^+ \rho_{4,4}-\gamma_{R}^{-} \rho_{7,7}+\Gamma_{2R}^+ \rho_{8,8}+J \text{Im}(\rho_{3,7}), \nonumber \\
&\text{Im} (\dot{\rho}_{3,7}) =J (\rho_{3,3}-\rho_{7,7})/2-\Gamma_D \text{Im} (\rho_{3,7}), \nonumber \\
&\text{Re} (\dot{\rho}_{3,7}) =-\Gamma_D \text{Re} (\rho_{3,7}),
\end{align}
with $\gamma_\nu^-=\Gamma_{1\nu}^{-} +\Gamma_{2\nu}^-$ and $\Gamma_D=(\gamma_L^-+\gamma_R^-)/2$; rates are denoted accordingly to the convention defined at the end of Sec.~\ref{sec:model}. As one can note, in the considered system evolution of the real part of the elements $\rho_{3,7}$ and $\rho_{7,3}$ is decoupled from dynamics of the other elements, and at the steady state it is equal to 0; therefore in the following part of the paper I take $\rho_{3,7}=-\rho_{7,3}=i \text{Im} (\rho_{3,7})$.

Then, I define the auxiliary system $A$ described by the density matrix $\hat{\rho}^A$ with matrix elements defined in the following way: $\rho^A_{i,j}=\rho_{i,j}/2$ for $i,j \in [1,9]$, $\rho^A_{i,j}=\rho_{i-9,j-9}/2$ for $i,j \in [10,18]$, $\rho^A_{3,16}=-\rho^A_{16,3}=\rho_{3,7}/2$ and $\rho^A_{12,7}=-\rho^A_{7,12}=\rho_{3,7}/2$, with all other elements equal to 0. Equations~\eqref{dynorig} can be then rewritten as
\begin{align} \label{dynaux}
&\dot{\rho}^A_{3,3}=\Gamma_{2L}^+ \rho^A_{2,2}-\gamma_{L}^{-} \rho_{3,3}^A+\Gamma_{1L}^+ \rho_{6,6}^A-J \text{Im}(\rho_{3,16}^A), \nonumber \\
&\dot{\rho}_{16,16}^A =\Gamma_{1R}^+ \rho_{13,13}^A-\gamma_{R}^{-} \rho_{16,16}^A+\Gamma_{2R}^+ \rho_{17,17}^A+J \text{Im}(\rho^A_{3,16}), \nonumber \\
&\text{Im} (\dot{\rho}_{3,16}^A) =J (\rho_{3,3}^A-\rho_{16,16}^A)/2-\Gamma_D \text{Im} (\rho_{3,16}^A), \nonumber \\ 
&\dot{\rho}^A_{12,12}=\Gamma_{2L}^+ \rho^A_{11,11}-\gamma_{L}^{-} \rho_{12,12}^A+\Gamma_{1L}^+ \rho_{15,15}^A-J \text{Im}(\rho_{12,7}^A), \nonumber \\
&\dot{\rho}_{7,7}^A =\Gamma_{1R}^+ \rho_{4,4}^A-\gamma_{R}^{-} \rho_{7,7}^A+\Gamma_{2R}^+ \rho_{8,8}^A+J \text{Im}(\rho^A_{12,7}), \nonumber \\
&\text{Im} (\dot{\rho}_{12,7}^A) =J (\rho_{12,12}^A-\rho_{7,7}^A)/2-\Gamma_D \text{Im} (\rho_{12,7}^A).
\end{align}
In this way, the coherent oscillations $|3 \rangle \leftrightarrow |7  \rangle$ (i.e. ${|\! \! \uparrow \downarrow \rangle \leftrightarrow |\! \! \downarrow \uparrow \rangle}$) in the original system are represented by a pair of two coherent transitions $|3 \rangle \leftrightarrow |16  \rangle$ and $|12 \rangle \leftrightarrow |7  \rangle$ in the auxiliary system. The dynamics of the density matrix $\hat{\rho}_A$ is schematically represented in Fig.~\ref{fig:map}~(a). This figure clearly illustrates, that the applied procedure involves the formal duplication: the auxiliary system may be interpreted as a two copies of the original system, with the coherent transition $|3 \rangle \leftrightarrow |7 \rangle$ replaced by transitions between the copies. Dynamics and thermodynamics of the original and the auxiliary system are completely equivalent. 

One may note, that the coherent oscillations between the spin states and the tunneling through the second dot in the auxiliary system are represented by transitions in the rows of the graphical model, whereas tunneling through the first dot by transitions in the columns (one should be aware that the rates $\Gamma_{1\nu}^-$ are also a part of the decoherence rate $\Gamma_D$). In this way, the auxiliary system may be considered as a bipartite one. Let us now use the following convention: each basis state $|i \rangle_A$ of the auxiliary system can be considered as a product of the basis states of the ``row subsystem'' $R$ and the ``column subsystem'' $C$: $|i \rangle_A=|k \rangle_R |l \rangle_C$. Here $k \in[1,3]$ denotes the row in which the state is placed in Fig.~\ref{fig:map}~(a), whereas $l\in[1,6]$ denotes the column; for example, $|13 \rangle_A={|2 \rangle_R |4 \rangle_C}$. 

Let us now calculate the quantum mutual information between the subsystems $R$ and $C$, defined as~\cite{nielsen2010}
\begin{align} \nonumber
I^A = &\text{Tr}[\hat{\rho}^A \ln(\hat{\rho}^A)]-\text{Tr}[\hat{\rho}^{R} \ln(\hat{\rho}^{R})]-\text{Tr}[\hat{\rho}^{C} \ln(\hat{\rho}^{C})].
\end{align}
Here $\hat{\rho}^{R}$ is the reduced matrix of the subsystem $R$ defined as a partial trace over the system $C$:
\begin{align}
&\hat{\rho}^{R}= \text{Tr}_C (\hat{\rho}^{A}) = \sum_{l=1}^6 {}_C \langle l |\hat{\rho}^{A} | l \rangle_C.
\end{align}
Analogously, the reduced matrix of the subsystem $C$ reads
\begin{align}
&\hat{\rho}^{C}= \text{Tr}_R (\hat{\rho}^{A})  = \sum_{k=1}^3 {}_R \langle k |\hat{\rho}^{A} | k \rangle_R.
\end{align}
The time derivative of the quantum mutual information can be written as
\begin{align} \label{ratemuta}
\dot{I}^A=\sum_{i,j} \frac{\partial I^A}{\partial \rho^A_{i, j}} \dot{\rho}_{i,j}^A=\sum_{i,j} a_{i,j} \dot{\rho}_{i,j}^A,
\end{align}
where the coefficients $a_{i,j}= \partial I^A/\partial \rho^A_{i, j}$ are functions of the density matrix elements $\rho_{i,j}^A$.

One may suppose that the information flow between the subsystems can be calculated in a way similar to the presented by Horowitz and Esposito~\cite{horowitz2014} for the case of classical systems, i.e. by separating contributions to $\dot{I}^A$ associated with the dynamics in rows and columns. This can be written as
\begin{align}
\dot{I}^A=\sum_{i,j} a_{i,j} \dot{\rho}_{i,j}^{A(R)}+\sum_{i,j} a_{i,j} \dot{\rho}_{i,j}^{A(C)},
\end{align}
There is, however, a question how to do that precisely. Let us make an educated guess: terms $\dot{\rho}_{i,j}^{A(C)}$ include only the elements associated with the tunneling through the first dot, i.e. containing the rates $\Gamma_{1\nu}^\pm$. This include also the decoherence terms in non-diagonal elements. For example, terms $\dot{\rho}^{A}_{3,3}$ and $\text{Im} (\dot{\rho}_{3,16}^A)$ [cf. Eq.~\eqref{dynaux}] can be separated as
\begin{align}
\dot{\rho}^{A(R)}_{3,3} &=\Gamma_{2L}^+ \rho^A_{2,2}-\Gamma_{2L}^{-} \rho_{3,3}^A-J \text{Im}(\rho_{3,16}^A), \nonumber \\
\dot{\rho}^{A(C)}_{3,3} &=-\Gamma_{1L}^{-} \rho_{3,3}^A+\Gamma_{1L}^+ \rho_{6,6}^A, \nonumber \\
\text{Im} (\dot{\rho}_{3,16}^{A(R)}) & =J (\rho_{3,3}^A-\rho_{16,16}^A)/2-\Gamma_{D2} \text{Im} (\rho_{3,16}), \nonumber \\
\text{Im} (\dot{\rho}_{3,16}^{A(C)}) & =-\Gamma_{D1} \text{Im} (\rho_{3,16}),
\end{align}
where $\Gamma_{Di}=(\Gamma_{iL}^-+\Gamma_{iR}^-)/2$. The expression
\begin{align}
\dot{I}^A_1=\sum_{i,j} a_{i,j} \dot{\rho}_{i,j}^{A(C)},
\end{align}
can be then interpreted as the rate of change of the mutual information due to dynamics of the first dot. Analogously, $\dot{I}^A_2=\dot{I}^A-\dot{I}^A_1$ is the rate of change of the mutual information due to dynamics of the second dot. 

One may also arrange the system in another way, with transitions in the second dot placed in columns [Fig.~\ref{fig:map}~(b)]; let us refer to this arrangement as the auxiliary system $B$ with the density matrix $\hat{\rho}^B$ (which is equal to $\hat{\rho}^A$; different index is used to distinguish arrangements). Repeating the aforementioned procedure, one can calculate the rate of change of the mutual information due to dynamics of the second dot in the auxiliary system $B$:
\begin{align}
\dot{I}^B_2=\sum_{i,j} b_{i,j} \dot{\rho}_{i,j}^{B(C)},
\end{align}
where, in analogy to Eq.~\eqref{ratemuta}, coefficients $b_{i,j}= \partial I^B/\partial \rho^B_{i, j}$ are functions of the elements of the density matrix $\hat{\rho}^{B}$. It will be later shown that the information flows calculated in systems $A$ and $B$ are not equivalent.

\begin{figure}
	\centering
	\subfloat{\includegraphics[width=0.97\linewidth]{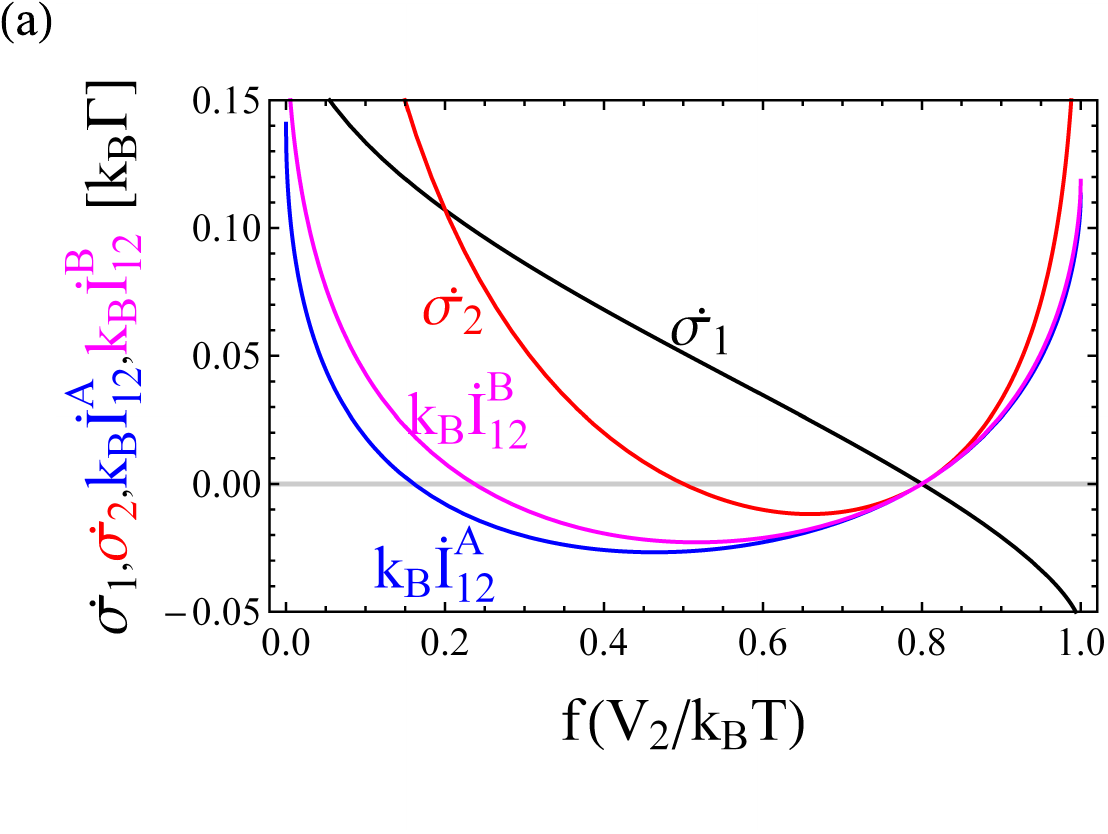}} \\
	\vspace{-5mm}
	\subfloat{\includegraphics[width=0.97\linewidth]{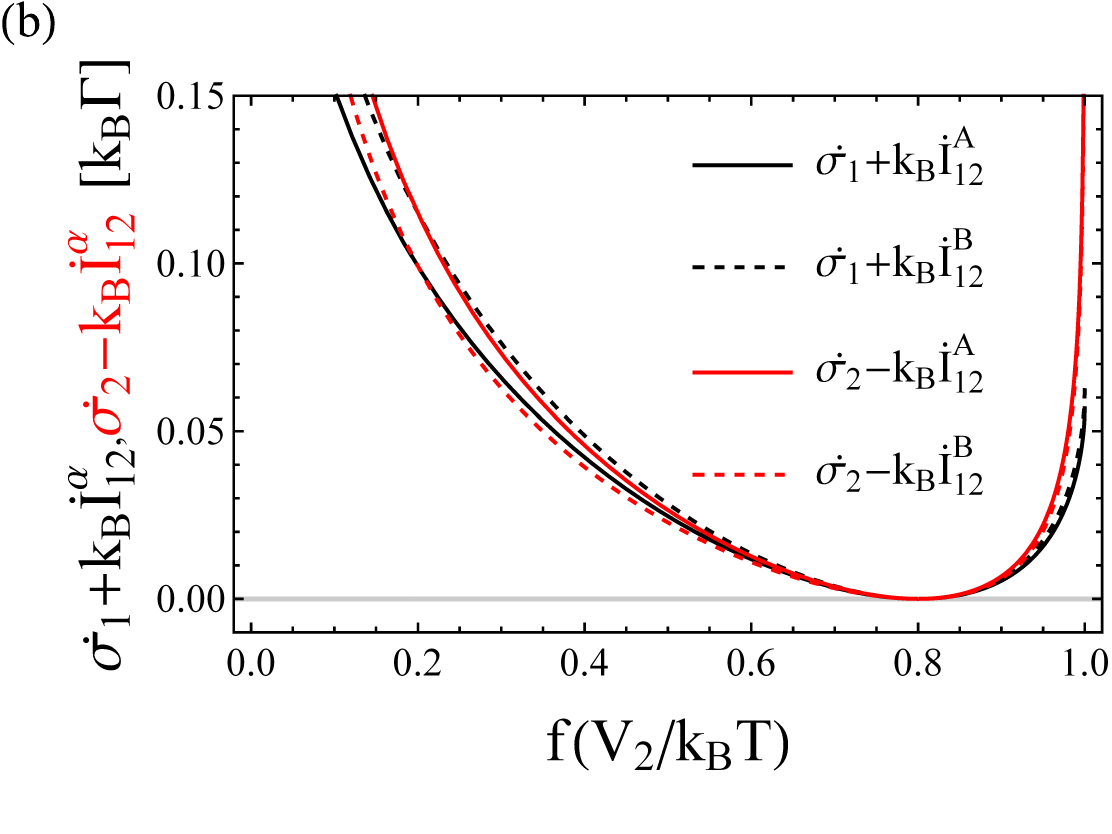}}
	\caption{Entropy production and the information flow as a function of $f(V_2 /k_B T)$ for $f(V_1 /k_B T)=0.2$, $\mu_{iL}=V_i/2$, $\mu_{iR}=-V_i/2$, $\epsilon_1=\epsilon_2=0$, $\Gamma_{1L}^\uparrow=\Gamma_{1R}^\downarrow=\Gamma_{2L}^\downarrow=\Gamma_{2R}^\uparrow=\Gamma$, $J=1.5\Gamma$.}
	\label{fig:wyk1}
\end{figure}

In the steady state $\dot{I}^\alpha=0$ (with $\alpha \in \{A, B\}$) and therefore $\dot{I}^\alpha_1=-\dot{I}^\alpha_2$. One may therefore define the information flow from the first to the second dot in different auxiliary systems as
\begin{align}
\dot{I}_{12}^A &=-\dot{I}^A_1, \\
\dot{I}_{12}^B &=\dot{I}^B_2.
\end{align}
Figure~\ref{fig:wyk1} presents the calculated entropy production rate and the information flow as a function of $f_2=f(V_2 /k_B T)$ for $f_1=f(V_1 /k_B T)=0.2$. As Fig.~\ref{fig:wyk1}~(a) shows, the entropy production rate in the second dot is negative for $f_2 \in (0.5,1-f_1)$, when the current in the second dot flows against the bias. In this range the information current $\dot{I}_{12}^\alpha$ is negative, since the first dot works as the demon which extracts information about the state of the second dot; therefore, the information is transferred from the second to the first dot. Conversely, for $f_2>1-f_1$ the second dot works as a demon (since the voltage at this dot is higher), the entropy production rate in the first dot is negative and the information flow is positive. 

One can also observe that the information flows calculated for auxiliary systems $A$ and $B$ are non-equivalent; therefore the mapping procedure is not unequivocal. One also finds that the quantum mutual information itself is not equivalent for different mappings: $I^A \neq I^B$ (not shown). This is associated with the fact that although dynamics of the whole system is the same for both mappings, separation into subsystems $R$ and $C$ is not. Physically, this nonequivalence of the separation into subsystems may be interpreted as a consequence of the fact that for different mappings the coherent transition ${|\! \! \uparrow \downarrow \rangle \leftrightarrow |\! \! \downarrow \uparrow \rangle}$ is in some way treated as the effective transition in either the first or the second dot (cf. Fig.~\ref{fig:map}). However, Fig.~\ref{fig:wyk1}~(b) shows that the local versions of the second law of thermodynamics which includes the information transfer, similar to the one derived by Horowitz and Esposito~\cite{horowitz2014} for classical bipartite systems, applies to the studied system for both mappings:
\begin{align}
\dot{\sigma}_1+k_B \dot{I}^\alpha_{12} \geq 0, \\
\dot{\sigma}_2-k_B \dot{I}^\alpha_{12} \geq 0.
\end{align}
Validity of these formulas was verified for different values of parameters. This shows that although the considered system is nonlocal due to presence of the quantum coherence between the first and the second dot one can describe the local thermodynamics of a single subsystem. It is worthwhile to note that a situation in which the information flow is not an unequivocally defined quantity is not unusual for Maxwell's demons. Similar situation in which different approaches give quantitatively different values of the information flow (which, however, satisfy the modified second law of thermodynamics) have been previously reported in Refs.~\cite{strasberg2014, barato2014, shirashi2016}.

\begin{figure}
	\includegraphics[width=0.97\linewidth]{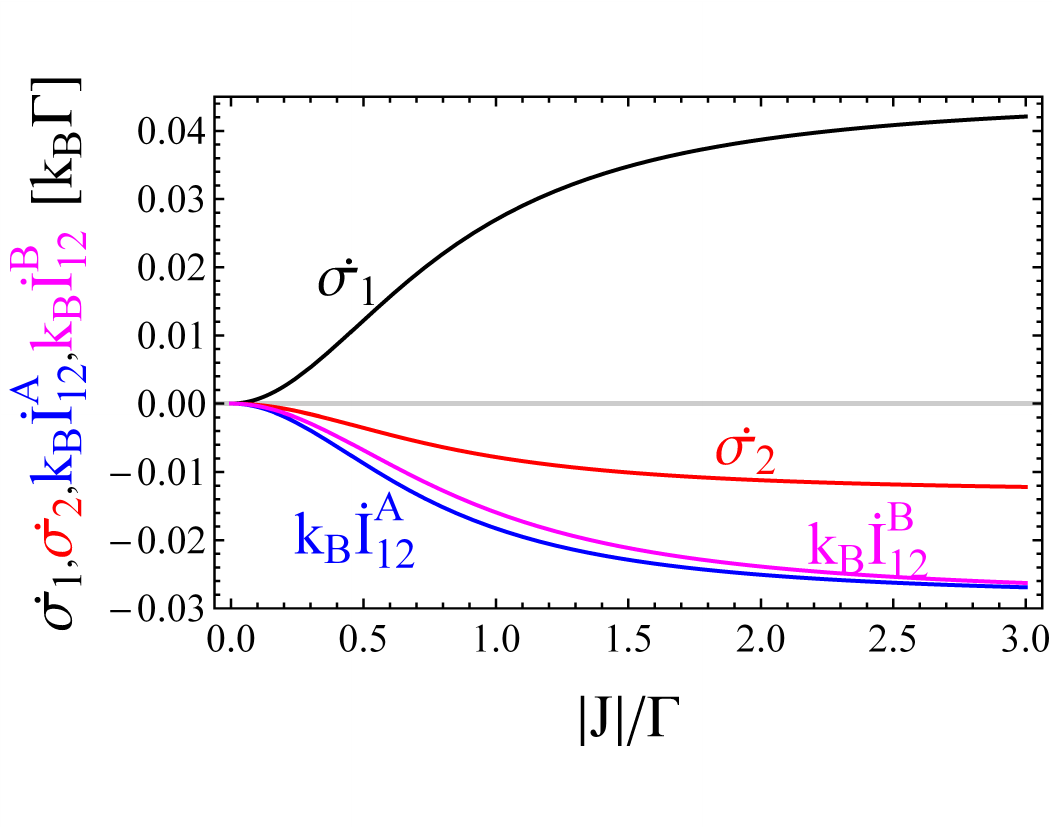}		
	\caption{Entropy production and the information flow as a function of $J$ for $\mu_{iL}=V_i/2$, $\mu_{iR}=-V_i/2$, $\epsilon_1=\epsilon_2=0$, $\Gamma_{1L}^\uparrow=\Gamma_{1R}^\downarrow=\Gamma_{2L}^\downarrow=\Gamma_{2R}^\uparrow=\Gamma$, $f(V_1 /k_B T)=0.2$, $f(V_2 /k_B T)=0.6$.}
	\label{fig:wyk2}
\end{figure}

Figure~\ref{fig:wyk2} presents the entropy production rate and the information flow as a function of $|J|$. As one may expect, power of the demon rises monotonically with the increased value of $|J|$; however, it is nearly saturated for relatively low values of $|J|$. One also observes that for $|J| \gg \Gamma$ the information flows calculated for different auxiliary systems become coincident. This is in line with the interpretation that nonequivalence of the information flow calculated for different mappings results from the fact that the coherent oscillations ${|\! \! \uparrow \downarrow \rangle \leftrightarrow |\! \! \downarrow \uparrow \rangle}$ are treated as effective transitions in one of the dots: For a high $|J|$ the timescale of the transition ${|\! \! \uparrow \downarrow \rangle \leftrightarrow |\! \! \downarrow \uparrow \rangle}$ is well separated from the timescale of the tunneling between the dots and the leads, and thus these processes do not compete with each other (i.e. decoherence of the spin states due to tunneling is much slower than the timescale of the coherent oscillations). As a result, it becomes equivalent whether one associates the transition ${|\! \! \uparrow \downarrow \rangle \leftrightarrow |\! \! \downarrow \uparrow \rangle}$ with the dynamics of the first or the second dot.

\section{Conclusions} \label{sec:conclusions}
I have studied the autonomous Maxwell's demon based on two exchange-coupled quantum dots, each attached to two fully spin-polarized leads in the anti-parallel configuration. The principle of operation of the demon is based on the coherent oscillations between the spin states of the system induced by the exchange interaction. The resulting dynamics can be described as the operation of the quantum iSWAP gate, due to which one of the dots acts as a feedback controller which reads out the spin state of the second dot and blocks transport with the bias while enabling tunneling in the reverse direction. This leads to the electron pumping against the bias, which generates the locally negative entropy production.

Moreover, the information transfer between the dots is described quantitatively by mapping the system onto the thermodynamically equivalent auxiliary one, which has the bipartite structure. This allows to separate contributions to the rate of change of the quantum mutual information associated with the dynamics of the first and the second dot, and thus define the information flow between the dots. Interestingly, one finds that in the considered system a sum of the entropy production in one dot and the information flow from this dot to another one is always nonnegative; this resembles the local version of the second law of thermodynamics derived by Horowitz and Esposito~\cite{horowitz2014} for the classical bipartite systems. The question, whether this result is an instance of some universal law, requires further studies. One may also consider how the approach used in this paper is related to the repeated interaction framework of Strasberg \textit{et al.}~\cite{strasberg2017}, which enabled a thermodynamic interpretation of some quantum master equations. 

\begin{acknowledgments}
I thank B. R. Bu\l{}ka for the careful reading of the manuscript and the valuable discussion. This work has been supported by the National Science Centre, Poland, under the Project No. 2016/21/B/ST3/02160.  
\end{acknowledgments}

\appendix*
%
\begin{figure}
	\centering
	\includegraphics[width=0.9\linewidth]{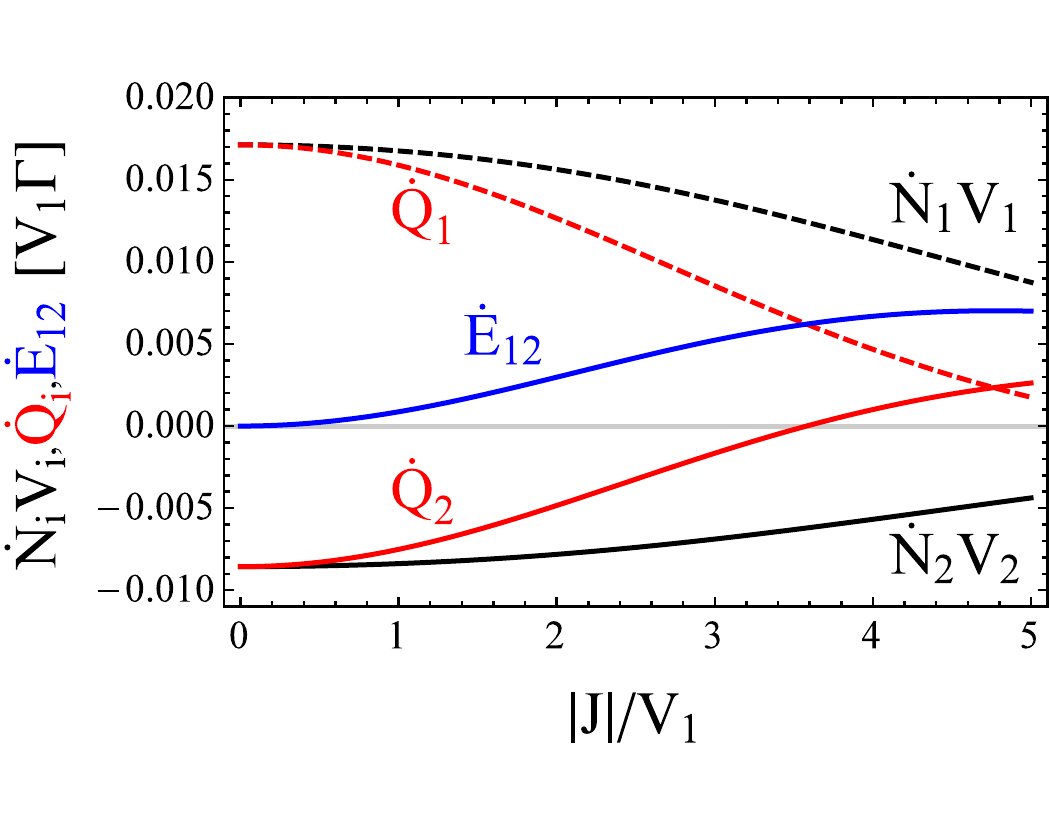}		
	\caption{Work inputs $\dot{N}_i V_i$, heat dissipated in the $i$th dot $\dot{Q}_i$ and the energy exchange between the first and the second dot $\dot{E}_{12}$ as a function of $|J|/V_1$ for $V_1=k_B T$ and $V_2=-k_B T/2$,  $\mu_{iL}=V_i/2$, $\mu_{iR}=-V_i/2$, $\epsilon_1=\epsilon_2=0$, $\Gamma_{1L}^\uparrow=\Gamma_{1R}^\downarrow=\Gamma_{2L}^\downarrow=\Gamma_{2R}^\uparrow=\Gamma$, $\Gamma_{1L}^\downarrow=\Gamma_{1R}^\uparrow=\Gamma_{2L}^\uparrow=\Gamma_{2R}^\downarrow=0$.}
	\label{fig:enflow}
\end{figure}
%

\section{Energy exchange between the dots}
In the main text it was assumed that for $|J| \ll V_i, k_B T$ the energy exchange between the dots can be neglected. Here I demonstrate the validity of this assumption. To achieve this goal, the transport in the system is described using the Pauli master equation~\cite{breuer2002, poltl2009}. Within this approach one uses the basis of the eigenstates of the Hamiltonian~\eqref{hamiltonian} instead of the basis of the localized states. In contrast to Eq.~\eqref{lindblad}, this method does not take into account the finite frequency of the coherent oscillations ${|\! \! \uparrow \downarrow \rangle \leftrightarrow |\! \! \downarrow \uparrow \rangle}$, and thus predicts non-vanishing current also for $J=0$. However, the given results may be considered as reliable for $|J|$ sufficiently higher than $\Gamma_{i \nu}^\sigma$. On the other hand, the method takes into account the finite level splitting caused by the exchange interaction. For $\Gamma_{i \nu}^\sigma \ll |J| \ll V_i, k_B T$ it gives the same results as Eq.~\eqref{lindblad}. The Pauli master equation reads
\begin{align}
\dot{\mathbf{p}}(t)=\mathbf{W} \mathbf{p}(t),
\end{align}
where $\mathbf{p}=(p_1, p_2, \dots)^T$ is the column vector of the eigenstate probabilities, whereas $\mathbf{W}$ is the rate matrix with the elements defined as
\begin{align}
W_{mn} =
\begin{cases}
\sum_{i, \nu, \sigma} (\Gamma^{i \nu \sigma+}_{mn}+ \Gamma^{i \nu \sigma-}_{mn}) &  \text{for} \quad m \neq n, \\
-\sum_{m \neq n} W_{mn}  &  \text{for} \quad m=n,
\end{cases}
\end{align}	
where
\begin{align} \nonumber
\Gamma_{mn}^{i\nu\sigma +}=\Gamma_{i\nu}^\sigma |\langle m |c^\dagger_{i \sigma}|n \rangle|^2 f[+ (E_m-E_n-\mu_{i\nu})/k_B T_{i \nu}], \\
\Gamma_{mn}^{i\nu\sigma -}=\Gamma_{i\nu}^\sigma |\langle m |c_{i \sigma}|n \rangle|^2 f[- (E_m-E_n-\mu_{i\nu})/k_B T_{i \nu}],
\end{align}
are the transition rates from the eigenstate $|n \rangle$ to the eigenstate $|m \rangle$ associated with the tunneling between the $i$th dot and the lead $i \nu$ (with the index $+/-$ denoting the tunneling to/from the dot). Here $E_m$ is the energy of the eigenstate $|m \rangle$. The steady state particle current through the $i$th dot can be calculated as
\begin{align}
\dot{N}_i = \sum_{n, m \neq n, \sigma} (\Gamma^{i L \sigma+}_{mn} - \Gamma^{i L \sigma-}_{mn}) p_n,
\end{align}
where $p_n$ is the steady state probability of the eigenstate $|n \rangle$, calculated by solving the equation $\mathbf{W} \mathbf{p}=0$. The heat dissipated in the $i$th dot reads~\cite{sanchez2012}
\begin{align}
\dot{Q}_i = (\mu_{i\nu}+E_m-E_n)\sum_{n, m \neq n, \nu, \sigma} (\Gamma^{i \nu \sigma+}_{mn} - \Gamma^{i \nu \sigma-}_{mn}) p_n.
\end{align}
Energy transfer from the first to the second dot $\dot{E}_{12}$ is a difference of the work input in the first dot $\dot{N}_1 V_1$ and the heat dissipated in the first dot $\dot{Q}_1$:
\begin{align}
\dot{E}_{12} = \dot{N}_1 V_1 - \dot{Q}_1=\dot{Q}_2-\dot{N}_2 V_2.
\end{align}

Figure~\ref{fig:enflow} shows the calculated work inputs $\dot{N}_i V_i$, dissipated heat and the energy exchange as a function of $|J|/V_1$ for $V_1=k_B T$ and $V_2=-k_B T/2$. It can be clearly seen that for $|J| \ll V_1$ the energy exchange tends to 0, whereas the heat dissipated in the second dot is finite and negative. Therefore the pumping against the bias is not a result of the energy exchange, but of the information flow. This justifies the assumption made in the main text. For a high ratio $|J|/V_1$ the current in the second dot is still pumped against the bias, but since the energy exchange becomes predominant the heat dissipated in the second dot starts to be positive and the systems ceases to work as a Maxwell's demon.

\end{document}